\documentclass[preprint,showpacs,preprintnumbers,amsmath,amssymb]{revtex4}

\newcommand\rf[1]{(\ref{eq:#1})}
\newcommand\lab[1]{\label{eq:#1}}

\newcommand\br{\begin{eqnarray}}
\newcommand\er{\end{eqnarray}}
\newcommand\be{\begin{equation}}
\newcommand\ee{\end{equation}}

\newcommand\lb{\lbrack}
\newcommand\rb{\rbrack}

\newcommand\bc{\begin{center}}
\newcommand\ec{\end{center}}

\newcommand\pa{\partial}

\newcommand{\ct}[1]{\cite{#1}}
\newcommand{\bib}[1]{\bibitem{#1}}
\newcommand\PRL[3]{\textsl{Phys. Rev. Lett.} \textbf{#1}, #3 (#2)}

\begin{document}

\preprint{arxiv:[hep-ph]}

\title{ Axion Photon Oscillations From a "Particle-Antiparticle" View Point  }

\author{E.I. Guendelman}%
\email{guendel@bgumail.bgu.ac.il}
% \author{E.I. Guendelman}%
% \email{guendel@bgumail.bgu.ac.il}
\affiliation{%
Department of Physics, Ben-Gurion University of the Negev \\
P.O.Box 653, IL-84105 ~Beer-Sheva, Israel
}%

\begin{abstract}
We observe that it is very usefull to introduce a complex field for 
the axion photon system in an external magnetic field, when for example considered with the 
geometry of the experiments exploring axion photon mixing, where the real part is the axion and 
the imaginary part is the photon polarization that couples to the axion when the magnetic field is present.
In the absence of the external magnetic field, the theory displays charge conjugation symmetry.
In this formulation the axion and photon are the symmetric and antisymmetric combinations of
particle and antiparticle (as defined from the complex field) respectively and they do not mix if 
the external magnetic field is set to zero. The magnetic field interaction is seen to be equivalent to first order
to the interaction of the complex charged field with an external electric potential, where this 
ficticious "electric potential" is proportional to the external magnetic field.
This interaction breaks the charge conjugation symmetry and therefore symmetric and antysymmetric
combinations are not mantained in time. As a result one obtains axion photon mixing in the
presence of an external magnetic field, a well known result understood in a different way.
\end{abstract}


\maketitle

%%%%%%%%%%%%%%%%%%%%%%%%%%%%%%%%%%%%%%%%%%%%%%%%%%%%%%%%%%%%%%%%%%%%%%%
\section{Introduction}

The idea of the axion \ct{weinberg} has been the subject of many investigations.
Although  was originally introduced in order to solve the strong CP problem, 
it has been postulated as a candidate for the dark matter also.
A great number of ideas and experiments for the search this particle have been proposed
\ct{Goldman}.

As for many pseudoscalar particles, for example among the already known particles
like the neutral pion, the axion field $\phi$ has coupling to the electromagnetism 
through an interaction term of the form 
$g \phi \epsilon^{\mu \nu \alpha \beta}F_{\mu \nu} F_{\alpha \beta}$.

A way to explore for observable consequences of the coupling of a light scalar 
to the photon in this way is to subject a beam of photons to a very strong magnetic field.

In this situation the axion photons system undergoes "oscillations".

This affects the optical properties of light which could lead to testable consequences\ct{PVLAS}. 
Also, the produced axions could be
responsible for the "light shining through a wall phenomena ", which are is obtained by first
producing axions out of photons through the oscillations obtained in a strong magnetic field region, then subjecting the mixed beam of photons
and axions to an absorbing wall for photons, but almost totally transparent to axions due to their weak 
interacting properties which can then
go through behind this "wall", applying then another magnetic field one can recover once again some photons 
from the produced axions \ct{LSTW}.

 In this article we will show that axion photon system allows a very natural complex structure,
and the theory aquires a very interesting form when one introduces a complex field that "unifies"
the axion and the photon. It is possible in fact to introduce a complex field for 
the axion photon system in an external magnetic field, when for example considered with the 
geometry of the experiments exploring axion photon mixing, where the real part is the axion and 
the imaginary part is the photon polarization that couples to the axion when the magnetic field is present.

In the absence of the external magnetic field, the theory displays charge conjugation symmetry.
In this formulation the axion and photon are the symmetric and antisymmetric combinations of
particle and antiparticle (as defined from the complex field) respectively and they do not mix if 
the external magnetic field is set to zero. The magnetic field interaction is seen to be equivalent to first order
to the interaction of the complex charged field with an external electric potential, where this 
ficticious "electric potential" is proportional to the external magnetic field.
This interaction breaks the charge conjugation symmetry and therefore symmetric and antysymmetric
combinations are not mantained in time. As a result one obtains axion photon mixing in the
presence of an external magnetic field, a well known result understood in a different way.

%%%%%%%%%%%%%%%%%%%%%%%%%%%%%%%%%%%%%%%%%%%%%%%%%%%%%%%%%%%%%%%%%%%%%%%%%%%%
\section{Action and Equations of Motion}
The action principle describing the relevant light pseudoscalar coupling to the photon is
\be
S =  \int d^{4}x 
\Bigl\lb -\frac{1}{4}F^{\mu\nu}F_{\mu\nu} + \frac{1}{2}\pa_{\mu}\phi \pa^{\mu}\phi - 
\frac{1}{2}m^{2}\phi^{2} + 
\frac{g}{2}g \phi \epsilon^{\mu \nu \alpha \beta}F_{\mu \nu} F_{\alpha \beta}\Bigr\rb
\lab{axion photon ac}
\ee

We now specialize to the case where we consider an electromagnetic field with propagation only along the z-direction
and where a strong magnetic field pointing in the x-direction is present. This field may have an arbitrary space dependence in z, but it is assumed to be time independent. In the case the magnetic field is constant, see for example \ct{Ansoldi} for general solutions.

For the small perturbations we consider only small quadratic terms in the action for the axion fields and the electromagnetic field, following the method of for example Ref. \ct{Ansoldi}, but now considering a static magnetic field pointing in the x direction  having
arbitrary z dependence and specializing to z dependent electromagnetic field perturbations and axion fields. This means that the interaction between the background field , the axion and photon fields
reduces to
 
\be
S_I =  \int d^{4}x 
\Bigl\lb \beta \phi E_x\Bigr\rb
\lab{axion photon int}
\ee

where $\beta = gB(z) $. Choosing the temporal gauge for the photon excitations and considering only the x-polarization for the electromagnetic waves, since only this polarization couples to the axion, we get the following 1+1 effective dimensional action
(A being the x-polarization of the photon)

\be
S_2 =  \int dzdt 
\Bigl\lb  \frac{1}{2}\pa_{\mu}A \pa^{\mu}A+ \frac{1}{2}\pa_{\mu}\phi \pa^{\mu}\phi - 
\frac{1}{2}m^{2}\phi^{2} + \beta \phi \pa_{t} A
\Bigr\rb
\lab{2 action}
\ee

($A=A(t,z)$, $\phi =\phi(t,z)$), which leads to the equations

\be
\pa_{\mu}\pa^{\mu}\phi + m^{2}\phi =  \beta \pa_{t} A
\lab{eq. ax}
\ee

\be
\pa_{\mu} \pa^{\mu}A = - \beta \pa_{t}\phi 
\lab{eq. photon}
\ee

As it is known, in temporal gauge, the action principle cannot reproduce the Gauss 
constraint (here with a charge density obtained from the axion photon coupling) and has
to be impossed as a complementary condition. However this 
constraint is automatically satisfied here just because of the type of dynamical reduction
employed and does not need to be considered  anymore.

\section{Introduction of the complex field, Charge Conjugation Symmetry and its Breakdown}
Without assuming any particular z-dependence for $\beta$, but still insisting that 
it will be static that the interaction term, after dropping  a total time derivative can
 be expressed as

\be
S_I =  \frac{1}{2} \int dzdt 
\beta \Bigl\lb \phi \pa_{t} A - A \pa_{t}\phi \Bigr\rb
\lab{axion photon int2}
\ee

defining the complex field $\psi$ as
\be
\psi = \frac{1}{\sqrt{2}}(\phi + iA)
\lab{axion photon complex}
\ee
we see that 
in terms of this complex field, the axion photon density takes the form
\be
S_I =  \frac{1}{2} \int dzdt 
\beta i( \psi^{*}\pa_{t}\psi - \psi \pa_{t} \psi^{*})
\lab{interaction complex}
\ee

We observe that to first order in $\beta$, \rf{interaction complex} that this interaction has the same 
form as that of scalar QED with an external "electric " field to first order.
In fact the magnetic field or more precisely $\beta /2$ appears to play the role of an "analog external electric potential". 

The free part of the action, that is the one containing kinetic and mass terms is in terms of the
complex field $\psi$

\be
S_{free} =  \int dzdt 
\Bigl\lb  \pa_{\mu}\psi^{*} \pa^{\mu}\psi - 
\frac{1}{4}m^{2}(\psi^{*} +\psi)^{2}  
\Bigr\rb
\lab{free action}
\ee

In the case the axion is taken to be massless an additional invariance, of phase change in the $\psi$
field appears. This leads to a conserved current and the theory becomes very close to scalar QED, something
that can be profitably exploited \ct{Guendelman} .
Now let us introduce the charge conjugation , that is,

\be
\psi \rightarrow \psi^{*}  
\lab{charge conjugation}
\ee

We see then that the free action \rf{free action} is indeed invariant under \rf{charge conjugation}.
The $A$ and $\phi$ fields when acting on the on the free vacuum give rise to a photon and an axion respectivelly,
but in terms of the particles and antiparticles defined in terms of  $\psi$, we see that a photon is an antisymmetric 
combination of particle and antiparticle and an axion a symmetric combination, since

\be
\phi =\frac{1}{\sqrt{2}}(\psi^{*} +\psi), A= \frac{1}{i\sqrt{2}}(\psi - \psi^{*})  
\lab{part, antipart}
\ee

So that the axion is even under charge conjugation, while the photon is odd.
These two eigenstates of charge conjugation will propagate without mixing as long as no external magnetic field is applied.
The interaction with the extenal magnetic field is not invariant under \rf{charge conjugation}, in fact
under \rf{charge conjugation} we can see that
\be
S_I \rightarrow - S_I
\lab{non invariance}
\ee

Therefore  these symmetric and antisymmetric combinations, corresponding to axion and photon are not going to be mantained in the presence on $B$  in the analog QED language, since the "analog external electric potential" breaks the symmetry between particle and antiparticle and therefore will not mantain in time the symmetric or antisymmetric combinations.  In fact if the analog external electric potential is taken to be a
repulsive potential for particles, it will be an attractive potential for antiparticles, so the symmetry breaking is maximal.

\section{Conclusions}
Pure axion and pure photon initial states correspond to symmetric and antisymmetric linear combinations of particle and antiparticle 
in the picture presented here. The reason these linear combinations are not going to be mantained in the presence on a non trivial $B$ in the analog QED language, is that the analog external electric potential breaks the symmetry between particle and antiparticle and therefore will not
mantain in time the symmetric or antisymmetric combinations.

%\vspace{-0.5cm}
\section*{Acknowledgments}

% \vspace{-0.5cm}
I would like to thank Stephen Adler for a great number of exchanges and for clarifications from his part on diverse aspects of the subject 
of axion-photon mixing experiments. I would also like to thank the Bulgarian Academy of Sciences for Hospitality.

%%%%%%%%%%%%%%%%%%%%%%%%%%%%%%%%%%%%%%%%%%%%%%%%%%%%%%%%%%%%%
% Doing references:                                         %
%%%%%%%%%%%%%%%%%%%%%%%%%%%%%%%%%%%%%%%%%%%%%%%%%%%%%%%%%%%%%

\end{document}